\documentclass{cernrep}

\begin{document}
\newcommand{\nc}{\newcommand}
\nc{\beq}{\begin{equation}}
\nc{\eeq}{\end{equation}}
\nc{\bea}{\begin{eqnarray}}
\nc{\eea}{\end{eqnarray}}
\nc{\nn}{\nonumber}
\nc{\pard}[2]{\frac{\partial#1}{\partial#2}}
\nc{\uq}{\hat{q}}
\nc{\un}{\hat{n}}
\nc{\R}{{\cal R}}
\nc{\Pow}{{\cal P}}

\title{Physics of the Cosmic Microwave Background and the Planck Mission}
\author{H. Kurki-Suonio}
\institute{Department of Physics, University of Helsinki, and Helsinki Institute of Physics, Finland}
\maketitle

\begin{abstract}
This lecture is a sketch of the physics of the cosmic microwave background.  The observed anisotropy can be divided into four main contributions: variations in the temperature and gravitational potential of the primordial plasma, Doppler effect from its motion, and a net red/blueshift the photons accumulate from traveling through evolving gravitational potentials on their way from the primordial plasma to here.
These variations are due to primordial perturbations, probably caused by quantum fluctuations in the very early universe. The ongoing Planck satellite mission to observe the cosmic microwave background is also described. 
\end{abstract}

\section{Introduction}

The cosmic microwave background (CMB) is radiation that comes from the early universe.
In the early universe, ordinary matter was in the form of hot hydrogen and helium plasma which was almost homogeneously distributed in space.  Almost all electrons were free.  Because of scattering from these electrons, the mean free path of photons was short compared to cosmological distance scales: the universe was opaque. As the universe expanded, this plasma cooled, and first the helium ions, then also the hydrogen ions captured the free electrons: the plasma was converted into gas and the universe became transparent.  After that the photons of the thermal radiation of this primordial plasma have travelled through the universe and we observe them today as the CMB. The CMB is close to isotropic, i.e., the microwave sky appears almost equally bright in every direction. With sensitive instruments we can, however, see small variations, the CMB anisotropy.

\begin{figure}[ht]
\begin{center}
\includegraphics[width=14cm]{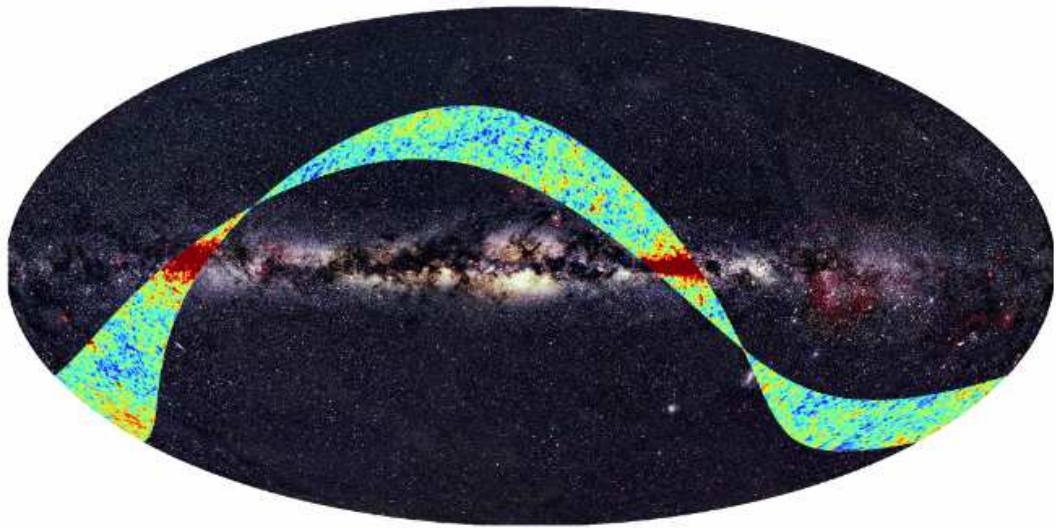}
\caption{The sky at optical and microwave wavelengths: A sky map of the first two week of observations by the Planck satellite at the 70 GHz frequency, superimposed on an optical image of the sky. From Ref.~\cite{Psite}. \emph{Credit: ESA, LFI \& HFI Consortia (Planck), Optical image: Axel Mellinger.}}
\label{fig:fls}
\end{center}
\end{figure}

There is also ``foreground'' microwave radiation that comes from astrophysical sources, our own galaxy and other galaxies. In Fig.~\ref{fig:fls} we see the radiation from the Milky Way as a horizontal red band in the microwave image, whereas further away from the galactic plane we see variations in the intensity of the CMB. The foreground can be separated from the CMB by measuring at several frequencies, since it has a different electromagnetic spectrum.

The formation of helium and hydrogen atoms is called recombination, although in this context it is a misnomer, since this was the first time the ions and electrons formed atoms.  The related increase of the photon free mean path beyond cosmological distance scales is called photon decoupling.  This happened when the age of the universe was about 380 000 years old. At this time there were small density variations, about one part in ten thousand in the primordial plasma/gas.  After photon decoupling, the over-densities began to grow by gravitational attraction and eventually led to the formation of galaxies and stars hundreds of millions of years later.

\begin{figure}[ht]
\begin{center}
\includegraphics[width=16cm]{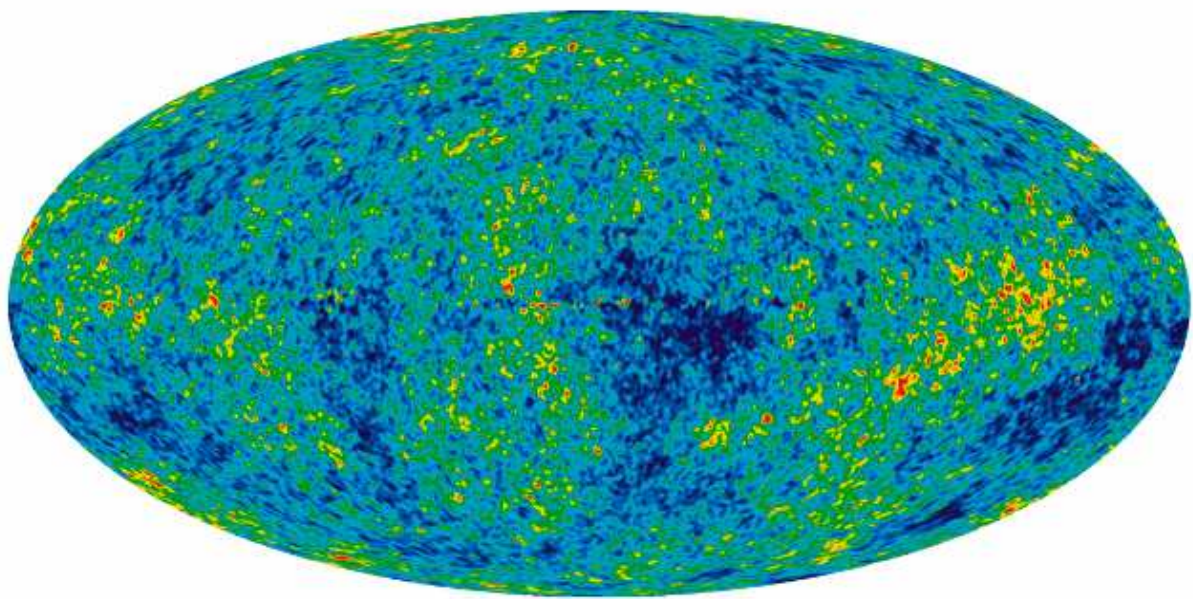}
%\vspace{-2cm}
\caption{Temperature anisotropy of the CMB according to 7 years of measurements by the WMAP satellite. This is a false-color image, where yellow and red indicate hotter than average, and blue colder than average. From Ref.~\cite{WMAPsite}. \emph{Credit: NASA / WMAP Science Team.}}
\label{fig:wmap7}
\end{center}
\end{figure}

When looking at the CMB we are thus looking at the 380 000 year old early universe.  We see those distant parts of the universe from which it has taken the whole remaining part of the history of the universe for the light to travel from there to here. The observed small variations in the CMB reflect the small density variations at that time. See Fig.~\ref{fig:wmap7}.

\begin{figure}[ht]
\begin{center}
\includegraphics[width=12cm]{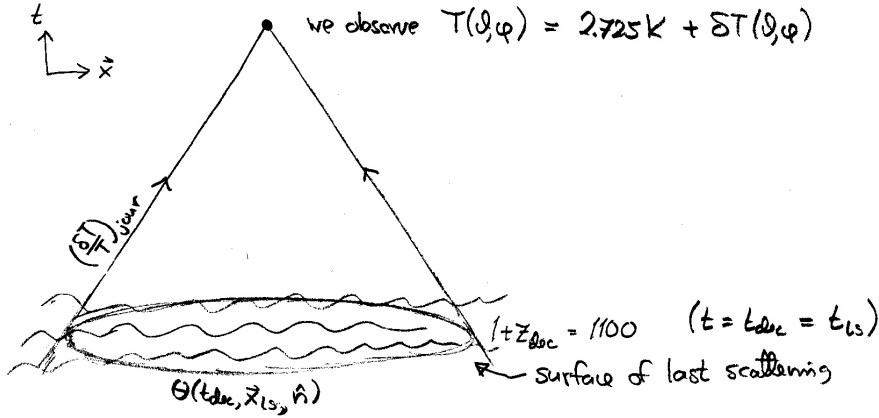}
\caption{A spacetime diagram of our past light cone.}
\label{fig:intr_and_jour}
\end{center}
\end{figure}

Because of the finite speed of light, everything we see lies on our past light cone 
(see Figure~\ref{fig:intr_and_jour}).  The intersection of our past light cone with the time of photon decoupling forms a sphere, which we call the sphere of last scattering. It is this sphere that we observe when we observe the CMB: we see each photon coming from the location where it last scattered from an electron. When the photons travel from the last scattering sphere to here they are redshifted by the expansion of the universe.  The universe has expanded by a factor of 1100 since last scattering, and therefore the photon wavelengths have been stretched by that factor. Photons decoupled when the temperature of the primordial plasma/gas was about 3000 K, and therefore the photons had then a blackbody spectrum with that temperature. When all wavelengths of a blackbody spectrum are stretched by the same factor, the spectrum remains blackbody, but its temperature falls with the same factor.  The observed mean temperature of the CMB is $T_0 = 2.725\pm0.001$ K today \cite{COBEtemp}.

However, because of the inhomogeneity of the universe, photons coming from different directions have suffered slightly different redshifts, which is another contribution to the observed CMB anisotropy. Thus the variation $\delta T(\theta,\phi)$ of the observed temperature $T(\theta,\phi) = 2.725 K + \delta T(\theta,\phi)$ can be divided into two contributions, $\delta T_\mathrm{intr}$ that is due to inhomogeneous conditions at the last scattering surface, and $\delta T_\mathrm{jour}$ that arises as the photons travel from the last scattering sphere to here.

An important thing of the anisotropy $\delta T(\theta,\phi)$ is that it is small.  The root-mean-square variation is  about $100 \mu$K, or
 \beq
  	\frac{\delta T}{T_0} \sim 4\times10^{-5} \,.
 \eeq
While this makes observing this anisotropy very difficult, it simplifies understanding and calculating the physics that causes this anisotropy: The primordial density perturbations $\delta\rho$ that are responsible for this anisotropy must have also been very small, and we can calculate their evolution using first-order perturbation theory around a homogeneous and isotropic model of the universe, the so-called background model.  The deviations from this background model are small, so
we can ignore any products of two or more such small quantities.  This makes the equations linear, so that they can be easily Fourier transformed, and lead to a direct relation between initial and final values.

CMB was discovered by Penzias and Wilson \cite{PeWi} in 1964, using a microwave antenna
at Bell Laboratories in Holmdel, New Jersey. The CMB anisotropy was first measured by the COBE satellite \cite{COBEanis} in 1992, and much more accurate measurements have later been taken by the WMAP satellite and are currently being taken by the Planck satellite.

In this lecture I sketch our present understanding how the CMB anisotropy arises. The relevant physics involves quantum field theory in curved spacetime (for the generation of primordial perturbation) and general relativistic perturbation theory (for their evolution and effect on the CMB), and it is not possible to give a full account in this short lecture. However, many parts of the relevant physics are relatively easy to understand, and 
I try to present those here; for the other parts I just have to give results without derivation, in an attempt to present a continuous story. 
I also give a short overview of the ongoing Planck satellite mission to observe the CMB.

\section{Background Universe}

The background model is the Friedmann-Robertson-Walker (FRW) universe, where the spacetime metric is
 \beq 
	ds^2 = -dt^2 + a(t)^2\left(dx^2+dy^2+dz^2\right) 
 \eeq
(we assume here the flat FRW universe, since it agrees well with observations, and simplifies the discussion). Here $a = a(t)$ is the scale factor that describes the expansion of the universe.  The coordinates $x$, $y$, $z$ are comoving coordinates that are related to actual distances via the scale factor.
The expansion rate is given by the Hubble parameter
 \beq
 	H(t) = \frac{1}{a}\frac{da}{dt}
 \eeq
whose present value $H_0 \equiv H(t_0)$ is called the Hubble constant ($t_0$ denotes the present time). The inverse of the Hubble parameter $H^{-1}$ is called the Hubble distance. It gives the distance over which causal effects can act in a cosmological time scale; in cosmology it is also called the \emph{horizon} scale. Today it is comparable to the size of the observable universe, but at early times it was much smaller than how far we can see today.

We denote the homogeneous quantities of the background model with an overbar, e.g., 
$\bar{\rho} = \bar{\rho}(t)$ is the energy density, $\bar{p} = \bar{p}(t)$ is the pressure, and so on.
The evolution of the background universe is governed by the Friedmann equations
 \bea
 	H^2 & = & \frac{8\pi G}{3}\bar{\rho}  \label{eq:Fried1} \\
	\frac{1}{a}\frac{d^2a}{dt^2} & = & -\frac{4\pi G}{3}(\bar{\rho}+3\bar{p})  \label{eq:Fried2}
 \eea
where $G$ is the gravitational constant.

There are at least four different kinds of energy and/or matter, whose energy density makes an important contribution to the universe: photons (the CMB), neutrinos, baryonic matter, and cold dark matter:
 \bea
 	\rho & = & \rho_\gamma + \rho_\nu + \rho_b + \rho_\mathrm{cdm} = \sum \rho_i \\
	p & = & p_\gamma + p_\nu + p_b + p_\mathrm{cdm} = \rho_\gamma/3 + \rho_\nu/3 + p_b = \sum p_i
 \eea
(Baryonic matter refers to ordinary matter made out of nucleons and electrons.) The early universe was radiation-dominated, most of the energy was in the form of photons, neutrinos (and even earlier, also other relativistic particles), so that $\bar{p} \sim \frac13\bar{\rho}$, leading to an expansion law $a \propto t^{1/2}$.  Later the universe became matter-dominated, so that $\bar{p} \ll \bar{\rho}$, leading to an expansion law $a \propto t^{2/3}$.  The transition from radiation domination to matter domination happened before photon decoupling.

It appears that a few billion years ago the expansion of the universe began to accelerate, implying that a fifth energy component, called dark energy, with negative pressure, $\bar{p} < -\frac13\bar{\rho}$, had become dominant.

\section{The Perturbed Universe}

In the FRW universe there are no stars or galaxies, and no anisotropy of the CMB, since it is completely homogeneous and isotropic.  The early part of the history of the universe, when deviations from homogeneity were small, and also large scales at later times, can be discussed using perturbation theory. The metric of this ``perturbed'' universe can be written as
 \beq 
	ds^2 = -(1+2\Phi)dt^2 + a^2(1-2\Psi)(dx^2+dy^2+dz^2)
 \label{eq:pert_metric}
 \eeq
where $\Phi(t,x,y,z)$ may be called the gravitational potential, since in the Newtonian limit of general relativity, it indeed becomes the Newtonian gravitational potential due to density perturbations: 
an over-density causes a negative $\Phi$, an under-density a positive $\Phi$.

The energy densities and pressures of the different components have perturbations,
 \beq
	\rho_i = (1+\delta_i)\bar{\rho_i} \qquad \mbox{and} \qquad p_i = \bar{p}_i + \delta p_i
 \eeq
Since the background model was isotropic, there was no net flow of energy, but in the perturbed universe we have also a mean velocity $\vec{v}_i$ of each energy component with respect to the coordinate system. The ``fluid'' perturbation variables $\delta_i$, $\delta p_i$, and $\vec{v}_i$ give a sufficient description of the perturbations in each component for as long as they can be approximated as perfect fluid, i.e., for as long as the mean free paths of their particles are short compared to the distance scales we are interested in.  In the perfect fluid approximation, the two metric perturbations become equal, $\Phi = \Psi$.  

After they have decoupled, the fluid description is not sufficient to describe the evolution of neutrinos and photons. Neutrinos decouple early, during the first second of the history of the universe. After that, the neutrino contribution causes a $\sim 10 \%$ difference between $\Phi$ and $\Psi$, until the universe becomes matter-dominated.

\section{Photon Distribution Function}

For the discussion of CMB anisotropy, we need a more detailed description of the photons, given by the photon distribution function $f(t,\vec{x},\vec{q})$, defined so that at a given time $t$ there are 
 \beq
 	dN = \frac{2}{(2\pi)^3}f(t,\vec{x},\vec{q})dVd^3q
 \eeq
photons with momentum in a range $d^3q$ around the value $\vec{q}$ within a volume $dV$ around location $\vec{x}$. (The factor 2 comes from the two polarization states of photons.) We divide the photon momentum $\vec{q}$ into its magnitude $q$ (photon energy) and direction $\hat{q}$ (a unit vector), $\vec{q} \equiv q\hat{q}$. 

In the background model, photons have the blackbody spectrum
 \beq  
 	\bar{f}(t,\vec{q}) = \frac{1}{e^{q/T(t)}-1}
 \label{eq:blackbody}
 \eeq
where $T(t)$ is the homogeneous temperature of the distribution.

In the perturbed universe we write 
 \beq
 	f = \bar{f} + \delta f \equiv \frac{1}{\exp\left\{\frac{q}{T(t)
	\left[1+\Theta(t,\vec{x},\vec{q})\right]}\right\}-1}
 \label{eq:fpert}
 \eeq
defining a momentum-dependent relative temperature perturbation $\Theta(t,\vec{x},\vec{q})$.  Any function $f(t,\vec{x},\vec{q})$ can be written in this form, but the important point is that to 1st order in perturbation theory, $\Theta$ does not develop any dependence on photon energy $q$.

Thus we have a radiation temperature perturbation called the \emph{brightness function}
 \beq
 	\Theta = \Theta(t,\vec{x},\hat{q})
 \eeq
which depends just on time, location, and photon direction.
Indeed, Fig.~\ref{fig:wmap7} is a plot of the measured CMB brightness function at our location and time as a function of the observation direction $-\hat{q}$. However, to predict its properties from theory, we need to follow how $\Theta(t,\vec{x},\hat{q})$ evolves from early times.

From Eqs.~(\ref{eq:blackbody}) and (\ref{eq:fpert}) the relation between $\delta f$ and $\Theta$ is
 \beq
 	\delta f = -q\pard{\bar{f}}{q}\Theta
 \label{eq:delf_theta}
 \eeq

We can integrate the brightness function over the direction angles to get various momenta, or \emph{multipoles}, of the photon distribution.  The three lowest ones are called the local monopole, dipole, and quadrupole of the photon perturbation:
 \bea
   	\Theta_0(t,\vec{x}) & \equiv & \frac{1}{4\pi}
		\int\Theta(t,\vec{x},\hat{q}) d\Omega \nn\\
   	\vec{\Theta}_1(t,\vec{x}) & \equiv & \frac{1}{4\pi}
		\int\hat{q}\Theta(t,\vec{x},\hat{q}) d\Omega \nn\\
 	\Theta_2^{ij}(t,\vec{x}) & \equiv & \frac{1}{4\pi}
		\int \left(\hat{q}^i\hat{q}^j-\frac13\delta_{ij}\right)
			\Theta(t,\vec{x},\hat{q}) d\Omega 
 \eea
(we denote the components of $\vec{x}$ and $\vec{q}$ by $x^i = (x^1,x^2,x^3)$ and $q^i = (q^1,q^2,q^3)$).
The monopole and dipole of the photon distribution give directly the photon density and velocity perturbation:
 \beq
 	\delta_\gamma = 4\Theta_0 \qquad\mbox{and}\qquad \vec{v}_\gamma = 3\vec{\Theta}_1
 \eeq

\section{Boltzmann Equation}
	
From statistical physics we have the Liouville theorem:  If there are now collisions between the particles, their distribution function will be constant along any particle trajectory in phase space:
 \beq 
 	\frac{df}{dt} \ \equiv \ \pard f t + \pard f{x^i}\frac{dx^i}{dt} + \pard f {q^i}\frac{dq^i}{dt} = 0 
 \eeq
(we sum over repeated indices),
where the total derivatives $dx^i/dt$ and $dq^i/dt$ refer to the motion of the particle.
Collisions between particles modify the equation by adding a \emph{collision term} $C[f]$, which depends on the distribution function:
 \beq
 	\frac{df}{dt} = C[f] \,.
 \eeq
Actually, it depends on the distribution functions of all species of particles the photons may collide with, in addition to the photon distribution function.

In curved spacetime, photons travel on lightlike geodesics. The photon coordinate velocity is given by
 \beq
 	\frac{dx^i}{dt} = \frac{\hat{q}^i}{a} \,.
 \eeq
The evolution of photon momentum and energy is obtained from the geodesic equation of general relativity, which can be derived from the metric. For the photon energy it gives
 \beq
 	\frac{dq}{dt} = q \left[ -H -\frac{\uq^i}{a}
\pard \Phi{x^i} + \pard\Psi{t}\right] \,.
 \eeq

To first order in perturbation theory
 \beq  \label{eq:q_approx}
 	\pard f {q^i}\frac{dq^i}{dt} \approx \pard fq \frac{dq}{dt} \,,
 \eeq
i.e., we care only about the dependence of the distribution function on, and the change of, photon energy, not photon direction.  This is because in the background model, the distribution function (\ref{eq:blackbody}) does not depend on the photon direction, and the photons do not change direction (except in collisions).  Thus both factors on the left-hand side of Eq.~(\ref{eq:q_approx}) have only a small ``first-order'' direction dependence, so that the direction dependence of the product is ``second-order small''.
Thus the photon Boltzmann equation becomes
 \beq
	\frac{df}{dt} = \pard f t + \frac{\uq^i}{a} \pard f{x^i} + q\pard fq 
	\left[ -H -\frac{\uq^i}{a} \pard \Phi{x^i} + \pard\Psi{t}\right] = C[f] \,.
 \label{eq:Boltz}
 \eeq 	

Of the five terms in Eq.~(\ref{eq:Boltz}), the first two are just kinematics: the distribution function changes in time, since photons move in and out of volume elements due to their velocity. The third term gives the change in photon energy (redshift) due to the overall expansion of the universe.
The two last terms give the effect of spacetime perturbations: the gravitational redshift due to a gradient in gravitational potential, and the effect of local variations in the expansion rate.

We can separate Eq.~(\ref{eq:Boltz}) into a background equation
 \beq
  	\frac{d\bar{f}}{dt} = \pard{\bar{f}}{t} - Hq\pard{\bar{f}}{q} = 0 
 \label{eq:bgfevol}
 \eeq
(the effect of collisions can be ignored at the background level)
and to a first-order perturbation equation
 \beq
	\frac{d(\delta f)}{dt} = \pard{(\delta f)}{t} + \frac{\uq^i}{a} \pard{(\delta f)}{x^i} 
	- Hq \pard{(\delta f)}{q}
	+ \pard{\bar{f}}{q}
	\left[ -H -\frac{\uq^i}{a} \pard \Phi{x^i} + \pard\Psi{t}\right] = C[f] \,. 
 \label{eq:fpertevol}
 \eeq
 	
From Eqs.~(\ref{eq:blackbody}) and (\ref{eq:bgfevol}) we obtain that the temperature of the background photon temperature falls as
 \beq
 	T \propto 1/a \,.
 \eeq
From Eqs.~(\ref{eq:delf_theta}) and (\ref{eq:fpertevol}) we obtain 
 \beq
 	\pard\Theta t + \frac{\uq^i}{a} \pard{\Theta}{x^i} 
	+ \frac{\uq^i}{a} \pard{\Phi}{x^i} - \pard\Psi t = C[\Theta] \,,
 \label{eq:brightness}
 \eeq
the \emph{brightness equation}.

\section{Thomson Scattering}

Photons scatter from charged particles.  At the time of interest, these are electrons, protons, and helium nuclei.  Since the scattering cross section is inversely proportional to the square of the mass of the charged particle, we need to consider just the electrons.  
In the non-relativistic (kinetic energies much below the electron mass) limit  scattering of photons on electrons is called Thomson scattering.  The differential cross section is
 \beq
 	\frac{d\sigma}{d\Omega} 
	= \frac{\sigma_T}{4\pi}\frac34\left(1+\cos^2\theta\right) \,,
 \label{eq:Thomson}
 \eeq
where
 \beq
 	\sigma_T \equiv \frac{8\pi}{3}\frac{\alpha^2}{m_e^2} = 6.65 \times 10^{-29}
	{\rm m}^2 \,.
 \eeq

The collision term Eq.~(\ref{eq:brightness}) is proportional to the electron density $n_e$, where only free electrons count, not those already bound in atoms. If the electron fluid were in rest ($\vec{v}_e = 0$), the effect of scattering would be to isotropize the photon distribution, i.e., to damp all its higher moments.  However, in the perturbed universe, there is a perturbation in the electron fluid velocity, equal to the baryon velocity perturbation, $\vec{v}_e = \vec{v}_b$, whose effect is to drag the photon velocity
perturbation towards it.  We skip the derivation of the collision term and just give the final form of the brightness equation:
 \beq
 	\pard\Theta t + \frac{\uq^i}{a} \pard{\Theta}{x^i} 
	+ \frac{\uq^i}{a} \pard{\Phi}{x^i} - \pard\Psi t 
	= n_e\sigma_T\left[\Theta_0 - \Theta(\uq) + \uq\cdot\vec{v}_b 
	+ \frac34\uq^i\uq^j\Theta_2^{ij}\right] \,.
 \label{eq:brightnesC}
 \eeq
The effect of the two first terms on the right-hand side (RHS) is to damp all multipoles of $\Theta(\uq)$, except $\Theta_0$.  The effect of the third term is to instead force $\vec{v}_\gamma = 3\vec{\Theta}_1$ towards $\vec{v}_b$.  The last term is due to the angular dependence of Eq.~(\ref{eq:Thomson}), which has a quadrupole shape, and has the effect that the quadrupole of the photon distribution is not damped as fast as the other multipoles.  

The differential cross section actually depends also on photon polarization (Eq.~(\ref{eq:Thomson}) is averaged over the two polarization directions). The quadrupolar angular dependence of this has the effect that $	\Theta_2^{ij}$ acts as a source of CMB polarization.  In this lecture, however, we discuss just the generation of the CMB temperature anisotropy, not its polarization.

\section{Line-of-Sight Integration}

The recombination of hydrogen had a dramatic effect on the brightness equation, since the density of free electrons $n_e$ dropped by many orders of magnitude.

Before recombination, $n_e$ was large, forcing the term in the brackets in Eq.~(\ref{eq:brightnesC}) to be very small.  We can then make the \emph{tight-coupling approximation}:
 \beq
 	\Theta(\uq) = \Theta_0 + \uq\cdot\vec{v}_b \quad\Rightarrow\quad \vec{v}_\gamma \equiv 3\vec{\Theta}_1 = \vec{v}_b;\quad \Theta_2^{ij} = 0
 \eeq

After recombination, $n_e$ was so small that most CMB photons have never scattered after recombination. We can then make the collisionless approximation, and use the \emph{collisionless brightness equation}
 \beq
 	\pard\Theta t + \frac{\uq^i}{a} \pard{\Theta}{x^i} 
	+ \frac{\uq^i}{a} \pard{\Phi}{x^i} - \pard\Psi t 
	= 0 \,.
 \label{eq:brightnes0}
 \eeq
The total derivative along a photon path can be written
 \beq
 	\frac{d}{dt} = \pard{ }{t} + \frac{\uq^i}{a}\pard{ }{x^i}
 \eeq
and thus we get from (\ref{eq:brightnes0}) that along a photon path
 \beq
 	\frac{d}{dt}\left(\Theta+\Phi\right) = \pard{\Phi}{t} + \pard{\Psi}{t} \,.
 \label{eq:effTevol}
 \eeq
The quantity $\Theta+\Phi$ is called the \emph{effective temperature perturbation}, since it adds to the local temperature perturbation the effect of the gravitational red/blueshift
from the local gravitational potential.

Although in reality hydrogen recombination lasted tens of thousands of years, and an exact calculation has to follow this, we can get a good qualitative understanding of the CMB anisotropy by making the \emph{instantaneous decoupling approximation}:
we assume that recombination took place at $t = t_\ast$ ($\approx 380\,000$ yr),
and use the tight-coupling approximation for $t < t_\ast$ and the collisionless approximation for $t > t_\ast$.

We can then integrate Eq.~(\ref{eq:effTevol}) along the photon path (\emph{line-of-sight integration}), starting at time $t_\ast$ from the location where the photon last scattered, $\vec{x}_{ls}$, to the present time $t_0$ and the location $\vec{x}_\mathrm{obs}$ where the photon is observed today:
 \bea  \label{eq:cmb_contr1}
	\Theta(t_0,\vec{x}_\mathrm{obs},\uq) + \Phi(t_0,\vec{x}_\mathrm{obs})    
 	& = & (\Theta+\Phi)(t_\ast,\vec{x}_{ls},\uq) 
	+ \int_{t_\ast}^{t_0}\left(\pard{\Phi}{t} + \pard{\Psi}{t}\right) dt \\
	& = & \Theta_0(t_\ast, \vec{x}_{ls}) + \Phi(t_\ast, \vec{x}_{ls})
	+ \uq\cdot\vec{v}_{b\gamma} + 
	\int_{t_\ast}^{t_0}\left(\pard{\Phi}{t} + \pard{\Psi}{t}\right) dt 
 \label{eq:cmb_contr2}
 \eea

Apply now this result to a fixed observer, looking at all directions.  The observed perturbation in the CMB temperature in direction $\un = (\theta,\phi)$ is given by
$\Theta(t_0,\vec{x}_\mathrm{obs},-\un)$, since the observer is looking against the photon direction.  The term $\Phi(t_0,\vec{x}_\mathrm{obs})$, which is just the gravitational potential of the observing site, does not depend on the direction looked at, and thus appears just an overall shift in the mean CMB temperature.
This effect is smaller than the accuracy the mean CMB temperature has been measured with, and we ignore it.  On the RHS we see four different contributions to the CMB temperature anisotropy:
\begin{itemize}
\item The original temperature perturbation at the last scattering sphere, $\Theta_0(t_\ast, \vec{x}_{ls}) = \frac14\delta_\gamma$
\item The gravitational potential from which the CMB photons have to climb (or fall) from, $\Phi(t_\ast, \vec{x}_{ls})$, causing a gravitational red/blueshift of the radiation temperature
\item A Doppler effect $\uq\cdot\vec{v}_{b\gamma} = -\un\cdot\vec{v}_{b\gamma}$ coming from the motion of the primordial baryon-photon fluid at the last scattering sphere
\item An effect that comes from the time dependence of the metric perturbations along the photon path. If the gravitational potential $\Phi$ does not depend on time, the redshift due to falling in it is canceled by the blueshift due to climbing from it.  Thus gravitational potential along the photon path has a net effect only if it is time dependent, so that this cancellation is not exact. The same applies to the perturbation $\Psi$ in the expansion rate. This effect is called the \emph{integrated Sachs-Wolfe (ISW) effect}.
\end{itemize}

The metric perturbations $\Phi$, $\Psi$ are affected by all energy components, 
$\rho_b$, $\rho_\mathrm{cdm}$, $\rho_\gamma$, $\rho_\nu$, and therefore we need the
evolution equations for all of them, in addition to the Einstein equations from general 
relativity for the evolution of $\Phi$, and $\Psi$. 
To obtain the quantities needed in Eq.~(\ref{eq:cmb_contr2}), we need to integrate these 
evolution equations starting from initial conditions specified at some time well 
before recombination, when the universe was still radiation dominated, and all scales of interest 
were ``outside the horizon'', meaning that the Hubble distance was then 
smaller than these scales.

These initial conditions for the perturbations are called \emph{primordial perturbations}, and to obtain a theoretical prediction for the observed CMB anisotropy (as well as for the observed matter distribution today), we need a theory for the production of primordial perturbations.

\section{Primordial Perturbations}

The primordial perturbations were apparently produced by some random process.  Therefore we only expect to predict their statistical properties.  The current favorite scenario for their production is called \emph{inflation}. Inflation refers to an accelerating expansion of the universe by a very large factor at very early times. Because the expansion is accelerating, the scale factor $a$ grows much faster than the Hubble distance, causing perturbations to exit the horizon (their distance scale becomes larger than the Hubble length).   During inflation microscopic scales were expanded to astronomical scales, and the primordial perturbations are produced from quantum fluctuations at these microscopic scales. 

There are many proposed theories where inflation can be realized, but in the simpler ones
there is only one dynamically important independent quantity at that time, a scalar field 
$\varphi$, called the \emph{inflaton}.  During inflation, the homogeneous background value
$\bar{\varphi}(t)$ ``rolls'' slowly towards the minimum of the inflaton potential $V(\varphi)$. All particles in the later universe are produced after inflation from the energy that was stored in the inflaton field during inflation, in a process called reheating. All perturbations arise from the inflaton perturbations $\delta\varphi$. Since perturbations in all quantities originate from a single perturbation quantity, they are related to each other in a simple manner, i.e., the resulting primordial perturbations are \emph{adiabatic}. 

This means, e.g., that the perturbations in the number densities of all particle species are the same
 \beq
 	\delta\left(\frac{n_i}{n_\gamma}\right) = 0 \qquad 
	\Rightarrow \quad \frac{\delta n_i}{n_i} = \frac{\delta n_\gamma}{n_\gamma} =
	\frac34\frac{\delta\rho_\gamma}{\rho_\gamma}
	= {\textstyle\frac34\delta_\gamma} \,.
 \eeq
(The photon number density is related to temperature by $n_\gamma \propto T^3$ and the photon energy density by $\rho_\gamma \propto T^4$.) 
For baryons and CDM, $\rho_i = m_in_i$, 
so that
 \beq
 	\delta_i = \frac{\delta\rho_i}{\rho_i} = \frac{\delta n_i}{n_i} = {\textstyle\frac34\delta_\gamma} \equiv \delta_m \,.
 \eeq

After inflation, all cosmological distances are much larger than the Hubble distance, and therefore perturbations at these scales do not have any dynamical evolution. 
These ``superhorizon'' perturbations are naturally described in terms of the associated spacetime curvature perturbation. We can define a ``comoving curvature perturbation'' $\R(t,\vec{x})$ (related to $\Phi$ and $\Psi$) that stays constant in time at superhorizon scales (for adiabatic perturbations). 

The quantum fluctuations during inflation are a random process, and therefore we can not predict individual values of $\R(\vec{x})$ from an inflation theory, but we can predict expectation values of the magnitudes of the perturbations at different distance scales.
These are given by the \emph{power spectrum}
 \beq
 	\Pow_\R(k) \equiv 
	\frac{{\cal V}}{2\pi^2}k^3\times\langle|\R_{\vec{k}}|^2\rangle
 \label{eq:Pow_def}
 \eeq
where $\langle\cdot\rangle$ denotes expectation value, and $\R_{\vec{k}}$ is the Fourier amplitude of $\R$ corresponding to wave vector $\vec{k}$.  Here ${\cal V}$ is the reference volume used to Fourier expand $\R(\vec{x})$, and its choice does not affect the result.
The inflation prediction for $\Pow_\R(k)$ is
 \beq
 	\Pow_\R(k) = \frac{1}{24\pi^2M_{Pl}^4}\frac{V(\varphi_k)}{\epsilon(\varphi_k)} \,,
 \label{eq:inf_Pow}
 \eeq
where $M_{Pl}$ is the Planck mass, $\varphi_k$ is the value of the inflaton field when scale $k$ exited the horizon ($k = aH$) during inflation, and
 \beq
 	\epsilon \equiv \frac{M_{Pl}^2}{2}\left(\frac{V'}{V}\right)^2 
	\ll 1 \,.
 \eeq

The primordial perturbations produced in inflation are close to scale invariant,
 \beq
 	\Pow_\R(k) \approx \mbox{const.} \,,
 \eeq
since during inflation $\varphi$ and $H$ change slowly, while the scale factor $a$ grows rapidly.
More accurately
 \beq
 	\Pow_\R(k) \approx A^2_s k^{n_s-1} \qquad
	\mbox{where}\qquad n_s-1 = -6\epsilon+2\eta
 \label{PAsns}
 \eeq
where $n_s$ is the \emph{spectral index} of the perturbations and
 \beq
 	\eta \equiv M_{Pl}^2\frac{V''}{V} \,, \qquad |\eta| \ll 1 \,.
 \eeq
For historical reasons, there is this $-1$ in the common definition of the spectral index $n_s$, but the relevant quantity is 
 \beq
 	n_s-1 \equiv \frac{d\ln \Pow_\R}{d\ln k} \,,
 \eeq
which gives the scale dependence of the primordial perturbations.
The \emph{slow-roll parameters} $\epsilon$ and $\eta$ depend on the inflation model, but they are always small in successful models. 
 
After inflation, as the universe gets older, the Hubble distance $H^{-1}$ grows, faster than the scale factor $a$, and encompasses larger scales.  At the photon decoupling time $t_\ast$, the Hubble distance was $\approx 200$ Mpc, corresponding to about $1^\circ$ on the CMB sky.  Thus at angles $\gg 1^\circ$ we see superhorizon perturbations that are still in their primordial state, i.e., they have not evolved since they were produced.

For these large scales, it is easy to derive an approximate prediction for the CMB anisotropy using Eqs.~(\ref{eq:cmb_contr2}) and (\ref{eq:inf_Pow}).  At $t_\ast$ the universe was already matter dominated by a factor of a few over radiation. Thus we approximate the total density perturbation $\delta \equiv \delta\rho/\rho$ by the matter density perturbation,
 \beq
 	\delta \approx \delta_m = {\textstyle \frac34}\delta_\gamma = 3\Theta_0
 \eeq
It turns out that $\Phi$ and $\Psi$ are constant in time in a matter-dominated universe.
Thus the ISW effect gets a contribution only from early times after photon decoupling, when the universe is not yet completely matter dominated, and at late times when the universe became dominated by dark energy. Therefore the last term of Eq.~(\ref{eq:cmb_contr2}) is subdominant and we ignore it in our approximation.
For adiabatic perturbations, velocity perturbations are negligible at superhorizon scales,
and thus we approximate Eq.~(\ref{eq:cmb_contr2}) by
 \beq
 	\frac{\delta T}{T}(\theta,\phi) 
	\approx {\textstyle\frac13}\delta + \Phi 
 \label{eq:cmb_large}
 \eeq
where the RHS refers to conditions on the last scattering sphere.

Now we still need to relate $\delta$, $\Phi$, and $\R$. In Newtonian gravity
 \beq
 	\nabla^2\Phi = 4\pi G\rho \,.
 \label{eq:Newtgrav}
 \eeq
Now our gravitational potential is due to density perturbations, so we replace
$\rho$ by $\delta\rho = \bar{\rho}\delta$. Since we are using comoving coordinates, we replace $\nabla^2$ by $(1/a^2)\nabla^2$. Using Eq.~(\ref{eq:Fried1}) we would then get
 \beq
 	\delta_{\vec{k}} = -\frac23\left(\frac{k}{aH}\right)^2\Phi_{\vec{k}}
 \eeq
in Fourier space.  The correct result derived from general relativity is
 \beq
 	\delta_{\vec{k}} = -\left[ 2 + \frac23\left(\frac{k}{aH}\right)^2\right]\Phi_{\vec{k}} \,,
 \eeq
showing that we get the Newtonian result for subhorizon scales ($k \ll aH$).  However, we are now discussing superhorizon scales, so we instead have the approximate result
 \beq
	\delta \approx -2\Phi
 \eeq
 
From general relativity, $\Phi$ and $\R$ are related by
 \beq
 	\Phi = -{\textstyle\frac35}\R
 \eeq
in a matter-dominated universe.

Thus the CMB anisotropy at large scales is
 \beq
  	\frac{\delta T}{T} \approx {\textstyle\frac13}\delta + \Phi 
	\approx -{\textstyle\frac23}\Phi + \Phi = {\textstyle\frac13}\Phi 
	= -{\textstyle\frac15}\R \,.
 \label{eq:cmb_R}
 \eeq
Before converting this result and Eq.~(\ref{eq:inf_Pow}) into a prediction of the statistical properties of CMB anisotropy, we need to discuss how the latter are described.

\section{CMB Angular Power Spectrum}

The observed CMB temperature variations form a function on a sphere (the celestial sphere). In general this refers just to the unit sphere of observation directions $\un$, but in the approximation (\ref{eq:cmb_large}) this corresponds to the last scattering sphere.  A standard way to analyze functions on a sphere is the expansion in terms of spherical harmonics
 \beq
	\frac{\delta T}{T}(\theta,\phi) 
	= \sum_{\ell m} a_{\ell m} Y_{\ell m}(\theta,\phi) \,, \qquad
	\ell = 0,1,\ldots \,, \quad m = -\ell,-\ell+1,\ldots,\ell \,,
 \eeq
where the harmonic coefficients $a_{\ell m}$ are obtained by
 \beq 
 	a_{\ell m} = \int d\Omega Y^\ast_{\ell m}(\theta,\phi) 
	\frac{\delta T}{T}(\theta,\phi)  \,.
 \eeq
This is analogous to the Fourier expansion of functions of three-dimensional space.
The different multipole numbers $\ell$ correspond to different angular scales, with a rough correspondence
 \beq
 	\theta \sim \frac{180^\circ}{\ell} = \frac{\pi}{\ell},.
 \label{eq:theta2ell}
 \eeq
The different $m$ for a given $\ell$ correspond to different patterns or orientations with the same angular scale.

The 	$a_{\ell m}$ depend linearly (through the linear physics of first order perturbation theory) on primordial perturbations.  
Since different Fourier modes evolve independently, their amplitudes are uncorrelated, and this lack of correlation is inherited by the multipole coefficients
 \beq
 	\langle a_{\ell m}a^\ast_{\ell' m'}\rangle = 0 
	\qquad\mbox{for}\qquad \ell \neq \ell' \mbox{ or } m\neq m' \,,
 \eeq
The evolution of the perturbations is different for different distance scales, but the physics does not differentiate between directions, and therefore Eq.~(\ref{eq:Pow_def}) depends just on the magnitude $k$ of $\vec{k}$.  The analogous property of the multipole coefficients is that the expectation values of their amplitudes depend on $\ell$ only, not on $m$.  This dependence is called the \emph{angular power spectrum}:
 \beq
	C_\ell \equiv \langle | a_{\ell m} |^2\rangle \,.
 \eeq

From this one obtains that the expectation value of the square of the temperature perturbation is given by a sum over the angular power spectrum,
 \beq
 	\bigg\langle\left(\frac{\delta T}{T}\right)^2\bigg\rangle 
	= \sum_\ell \frac{2\ell+1}{4\pi}C_\ell 
 \eeq
 
\begin{figure}[ht]
\begin{center}
\includegraphics[width=8cm]{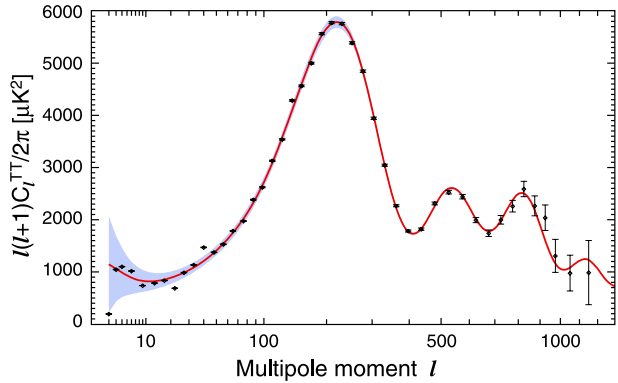}
\caption{The CMB angular power spectrum from 7 years of measurements by the WMAP satellite. Black dots with error bars represent the measurements and the red curve is 
a theoretical prediction from a best-fit cosmological model. The blue band represents the expected scatter (\emph{cosmic variance}) due to the randomness of the generation of primordial perturbations. From Refs.~\cite{lambda,WMAP7spec}. \emph{Credit: WMAP Science Team}.}
\label{fig:wmap7TT}
\end{center}
\end{figure}

Fig.~\ref{fig:wmap7TT} shows the observed CMB angular power spectrum based on 7 years of measurements by the WMAP satellite \cite{WMAP7paper}. The strong peak near $\ell \sim 200$ corresponds to structure at $1^\circ$ scale, which is prominent in Fig.~\ref{fig:wmap}.

\section{Large Scales}

We have a prediction for the three-dimensional power spectrum of primordial curvature
perturbations, Eq.~(\ref{eq:inf_Pow}), from inflation.  On the other hand, Eq.~(\ref{eq:cmb_R}) relates CMB anisotropy to the values of 
 \beq
 	\R(\vec{x}_{ls}) = \sum_{\vec{k}} \R_{\vec{k}} e^{i\vec{k}\cdot\vec{x}_{ls}}
 \eeq
on the last scattering sphere.  To get from an expansion in terms of plane waves to a spherical harmonic expansion, we need the relation between 3-dimensional plane waves and spherical harmonics
 \beq
 	e^{i\vec{k}\cdot\vec{x}_{ls}} =  
	4\pi\sum_{\ell'm'} i^\ell j_\ell(kx_{ls})Y_{\ell'm'}(\hat{x})
	Y^\ast_{\ell'm'}(\hat{k}) \,.
 \eeq
Here $x_{ls}$ is the coordinate distance to the last scattering sphere, and $j_\ell$ are spherical Bessel functions.

\begin{figure}[ht]
\begin{center}
\includegraphics[width=6cm]{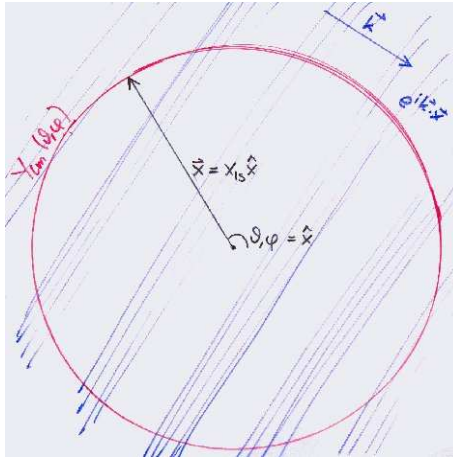}
\caption{A plane wave intersecting the last scattering sphere.}
\label{fig:plane_wave_on_sphere}
\end{center}
\end{figure}

Fig.~\ref{fig:plane_wave_on_sphere} illustrates how a Fourier mode (a plane wave) contributes to different angular scales at different parts of the last scattering sphere.  Thus a given wave number $k$ contributes to many multipoles $\ell$, as given by
$j_\ell(kx_{ls})$, but the maximum contribution is around
 \beq
 	\ell \sim kx_{ls} \,.
 \eeq

Now it is straightforward to calculate
 \beq
 	C_\ell \equiv \frac{1}{2\ell+1}\sum_m\langle|a_{\ell m}|^2\rangle = \ldots 
 	= \frac{4\pi}{25}\sum_{\vec{k}} \langle|\R_{\vec{k}}|^2\rangle j_\ell(kx)^2\
	= \frac{4\pi}{25}\int\frac{dk}{k}\Pow_\R(k)j_\ell(kx)^2 \,.
 \eeq
For the case of a scale-invariant power spectrum, $\Pow_\R = \mbox{const.}$, we can use the property
 \beq
 	\int\frac{dk}{k}j_\ell(kx)^2 = \frac{1}{2\ell(\ell+1)}
 \eeq
of spherical Bessel functions to arrive at
 \beq
 	C_\ell = \frac{\Pow_\R}{25}\cdot\frac{2\pi}{\ell(\ell+1)} \,,
 \eeq
or, using Eq.~(\ref{eq:inf_Pow}),
 \beq
 	\frac{\ell(\ell+1)}{2\pi}C_\ell 
	= \frac{\Pow_\R}{25} = \frac{1}{600\pi^2M_{Pl}^4}\frac{V}{\epsilon} \,.
 \label{eq:result}
 \eeq
This result is the reason why the CMB angular power spectrum is customarily plotted as
$(\ell(\ell+1)/(2\pi))C_\ell$.  

Eq.~(\ref{eq:result}) should apply for large scales, $\theta \gg 1^\circ$, 
or $\ell \ll 180$. From Fig.~\ref{fig:wmap7TT} we read that the observed value is about $1000\mu{\rm K}^2/T_0^2 \approx 1.3\times10^{-10}$.  This gives a constraint for inflation models
 \beq
 	\frac{V(\varphi_x)}{\epsilon(\varphi_x)} 
	\approx 8\times10^{-7} M_{Pl}^4\approx (0.03 M_{Pl})^4 
 \eeq
where $\varphi_x$ refers to the value of the inflaton field when the observed cosmological scales exited the horizon.  Since $\epsilon \ll 1$, this gives an upper limit to the inflation energy scale
 \beq
 	V(\varphi_x)^{1/4} < 0.03 M_{Pl} = 7\times10^{16}\mbox{ GeV} \,.
 \eeq

\section{Smaller Scales}

For smaller scales, $\theta < 1^\circ$, or $\ell > 180$, the perturbations enter the horizon before $t_\ast$, and therefore they have had time to evolve from their primordial state before we observe them. We give only a qualitative discussion of the main features of the relevant physics. The gravitational attraction of the overdense regions causes the tightly coupled photon-baryon fluid to fall into their gravitational wells. However, this increases the radiation pressure, which eventually pushes it out.  Thus the fluid begins to oscillate, moving in and out of the gravitational wells. The gravitational potential is dominated by cold dark matter, which does not feel the radiation pressure, and therefore does not participate in these \emph{acoustic oscillations}.  Different Fourier modes of the perturbations oscillate with different frequencies, the relation between the wave number $k$ and (angular) frequency $\omega = c_sk$ given by the sound speed $c_s$ in the baryon-photon fluid,  
 \beq
 	c_s^2 = \frac13\frac{1}{1+R} \qquad\mbox{where}\qquad 
	R\equiv\frac34\frac{\bar{\rho}_b}{\bar{\rho}_\gamma}
 \eeq

The oscillation in the density of the baryon-photon fluid (or the temperature perturbation $\Theta_0$ proportional to it), is not symmetric, since it is biased by the gravitational potential of the CDM: the maximum over-densities in the potential wells are always larger than the under-densities at the opposite phase of the oscillation. A proper calculation gives that the quantity that oscillates around zero is proportional to $\Theta_0 + (1+R)\Phi$, so that we have
 \beq
 	\Theta_{0\vec{k}} + (1+R)\Phi_{\vec{k}} \propto \cos c_skt
 \label{eq:osc}
 \eeq
if we ignore the expansion of the universe.  The expansion causes $R$ and $c_s$ to change with time, so that we have to replace $c_st$ in Eq.~(\ref{eq:osc}) by the
\emph{sound horizon}
 \beq
 	r_s(t) \equiv \int_0^t \frac{c_s(t')}{a(t')}dt' \,,
 \eeq
the coordinate distance travelled at sound speed by time $t$.

The first two terms in Eq.~(\ref{eq:cmb_contr2}) are thus given by
 \beq
 	(\Theta_0+\Phi)_{\vec{k}}(t_\ast) 
	\approx -R\Phi_{\vec{k}}(t_\ast) + A_{\vec{k}}\cos kr_s(t_\ast) \,.
 \label{eq:mono}
 \eeq
The amplitude $A_{\vec{k}}$ is complicated to derive 
from $\R_{\vec{k}}$, since it is affected by physics near the time of horizon entry, 
where neither superhorizon, nor subhorizon approximations apply; 
but for smaller scales it gets a notable boost by a gravitational driving effect due to the evolution of the gravitational potential $\Phi$ when the universe was not yet matter dominated: The baryon-photon fluid falls into an evolving gravitational well that is becoming weaker.  Thus when it is kicked out by the radiation over-pressure, it flies out further than from where it came, boosting the oscillation amplitude.

The most conspicuous feature of Eq.~(\ref{eq:mono}) is its oscillatory dependence on scale $k$:  It is maximal at those scales $k$, given by
 \beq
 	kr_s = m\pi \,, \qquad m = 1,2,\ldots
 \eeq
where the oscillations were just at their extrema when photons decoupled.  Thus we see a strong CMB anisotropy at the corresponding angular scales
 \beq
 	\theta \sim \frac{\pi}{kx_{ls}} \qquad\mbox{or}\qquad \ell \sim kx_{ls}
	=  m\pi\frac{x_{ls}}{r_s(t_\ast)} \equiv m\ell_A
 \eeq
where 
 \beq
 	\ell_A \equiv \pi\frac{x_{ls}}{r_s(t_\ast)} \equiv \frac{\pi}{\theta_s}
 \eeq
is called the \emph{acoustic scale in multipole space} and 
 \beq
  	\theta_s \equiv \frac{r_s(t_\ast)}{x_{ls}}
 \eeq
is the \emph{sound horizon angle}.

This phenomenon is responsible for the oscillatory behavior in the $C_\ell$ seen in 
Fig.~\ref{fig:wmap7TT}.  We get from the separation of the peaks the observed value for $\ell_A \sim 300$, which is a tight constraint on cosmological parameters which $r_s(t_\ast)$ and $x_{ls}$ depend on.

\section{Diffusion Damping}

\begin{figure}[ht]
\begin{center}
\vspace{0.9cm}   % Otherwise covers the section title
\includegraphics[width=10cm]{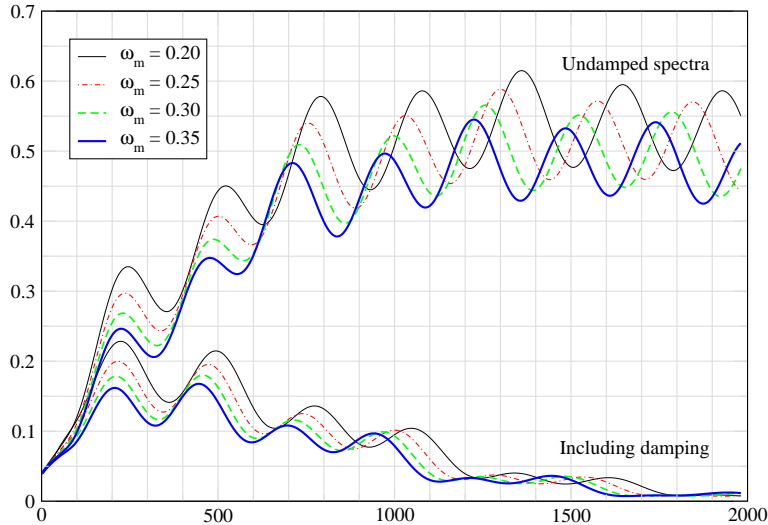}
\caption{The effective temperature, $\Theta_0+\Phi$, contribution to the angular power spectrum $C_\ell$, calculated both with
and without the effect of diffusion damping. The spectrum is given
for four different values of total matter density $\omega_m$, with baryonic matter density $\omega_b = 0.01$.
Figure and calculation by R.~Keskitalo. From Ref.~\cite{Keskitalo}.}
\label{fig:damping}
\end{center}
\end{figure}

The most important effect that we neglected in making the instantaneous decoupling approximation is \emph{photon diffusion}.  During recombination the photon mean free path grows rapidly. While the photons are still scattering, the photons carry energy from one part of the fluid to another, and this effect acts towards homogenizing the fluid temperature over a distance scale related to the photon mean free path.  This damps the temperature perturbations at the smaller scales. The effect on $C_\ell$ is quite dramatic as can be seen in Fig.~\ref{fig:damping}.

\section{Putting It All Together}

\begin{figure}[ht]
\begin{center}
\includegraphics[width=14cm]{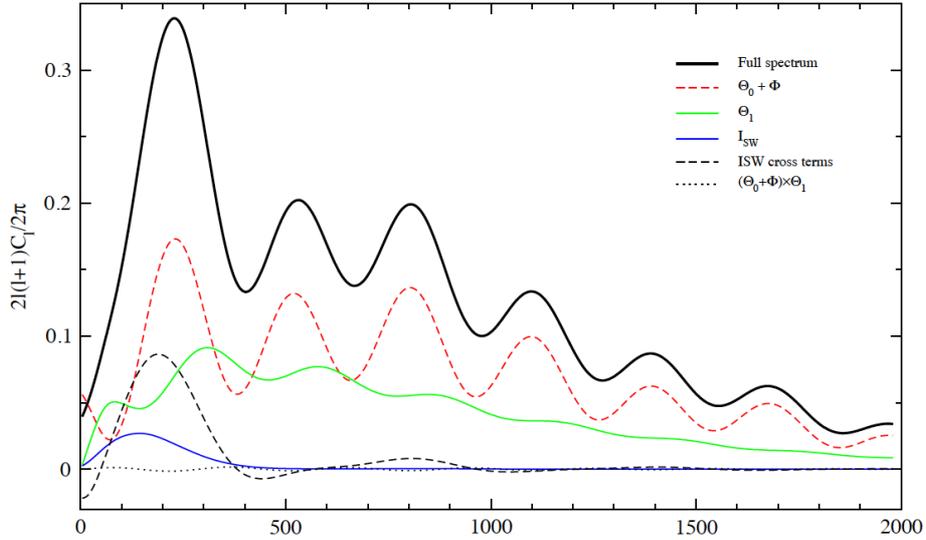}
\caption{The full $C_\ell$ spectrum calculated for the cosmological model
$\omega_m = 0.2$, $\omega_b =
0.03$, no dark energy, and $n_s=1$, and the different contributions to it. (The
calculation involves some approximations which allow the description
of $C_\ell$ as just a sum of these contributions and is not accurate enough for estimating cosmological parameters from data.) Here $\Theta_1$ denotes
the Doppler effect. Figure and calculation by R.~Keskitalo. From Ref.~\cite{Keskitalo}.}
\label{fig:spectrum_and_components_TT}
\end{center}
\end{figure}

The effective temperature perturbation $\Theta_0+\Phi$ is the dominant contribution to $C_\ell$, but the other two terms in Eq.~(\ref{eq:cmb_contr2}) are important also. In acoustic oscillation the fluid velocity $v_{b\gamma}$ oscillates too, but in a different phase: for a given Fourier mode, when the density is at the extremum, the fluid is momentarily at rest.  Thus the contribution from the Doppler effect $-\un\cdot\vec{v}_{b\gamma}$ to $C_\ell$ is also peaked, but at different $\ell$, and therefore it acts to partially fill the troughs between the peaks coming from $\Theta_0+\Phi$.

The $C_\ell$ is quadratic in $\delta T/T$ and therefore it also picks contributions from products of the three separate contributions $\Theta_0+\Phi$, $-\un\cdot\vec{v}_{b\gamma}$, and 
$\int (\partial\Phi/\partial t + \partial\Psi/\partial t)$.  Since the first two oscillate in different phases, their cross-contribution is small, but the ISW effect, which alone is the smallest of the three, is correlated with the effective temperature perturbation, and therefore actually contributes more through its cross-term with $\Theta+\Phi$, see Fig.~\ref{fig:spectrum_and_components_TT}.

\section{Cosmological Parameters}

We can compare the observed $C_\ell$ to those predicted by different cosmological models to determine the values of the free parameters in those models. Other cosmological data can usefully supplement CMB observations, but accurate CMB observations are vital for this task.  

The simplest cosmological model that fits current data is the $\Lambda$CDM model, which has:
\begin{itemize}
\item a flat background universe
\item primordial density perturbations with a constant spectral index,
\item which are adiabatic
\item no primordial gravitational waves
\item dark energy in the form of a cosmological constant $\Lambda$ (vacuum energy with a constant density)
\item negligible ($\ll 1$ eV) neutrino masses
\end{itemize}

This model has 6 parameters: the amplitude $A_s$ and spectral index $n_s$ of primordial perturbations (see Eq.~\ref{PAsns}); the background densities of baryonic matter $\omega_b$, cold dark matter $\omega_\mathrm{cdm}$, and the density of vacuum energy $\Omega_\Lambda$; and the optical depth of the universe after recombination $\tau$.
(In Figs.~\ref{fig:damping} and \ref{fig:spectrum_and_components_TT}, $\omega_m \equiv
\omega_\mathrm{cdm}+\omega_b$.)  

Here the densities are given in terms of density parameters
 \beq
 	\Omega_i \equiv \frac{\bar{\rho}_{i0}}{\rho_c}
 \eeq
where 
 \beq
 	\rho_c \equiv \frac{8\pi G}{H_0^2} = 1.88\times10^{-26}h^2\mbox{kg/m}^3
 \eeq
is the critical density for the universe required to make the background universe flat,
and $h \equiv H_0/(100\mbox{km/s/Mpc})$.  For baryonic and cold dark matter we have further defined $\omega_i \equiv \Omega_ih^2$, which give the average density of these components in the universe today as
 \beq
 	\bar{\rho}_{i0} = \omega_i\times1.88\times10^{-26}\mbox{kg/m}^3 \,.
 \eeq
The optical depth gives the expectation number of scatterings per CMB photon after recombination.  These are mainly due to free electrons that were liberated when the radiation from the first stars reionized the interstellar gas.

\section{WMAP Results}
\label{sec:wmap}

The WMAP satellite \cite{WMAPsite} began observing in August 2001, and ceased operations nine years later. WMAP measured the microwave sky at 5 frequency bands centered from 23 to 94 GHz \cite{WMAP7calib}.  To improve their sensitivity, the WMAP instruments were passively cooled with a solar shade and radiators to an operating temperature of 90 K.

The WMAP team has so far published results based on the first 7 years of data.
According to them \cite{WMAP7cosmo}, the values of the $\Lambda$CDM model parameters are
 \bea
 	A_s & = & 4.94\pm0.05\times10^{-5} \nn\\
	n_s-1 & = & -0.037\pm0.012 \nn\\
	\omega_b & = & 0.0226\pm0.0005 \nn\\
	\omega_\mathrm{cdm} & = & 0.112\pm0.004 \nn\\
	\Omega_\Lambda & = & 0.728\pm0.015 \nn\\
	\tau & = & 0.087 \pm0.014
 \eea
Except for the spectral index $n_s-1$ and optical depth $\tau$, these are already quite accurate.  However, we get this accuracy only by assuming the $\Lambda$CDM model.  If we relax these assumptions, we can use the data to derive upper limits on deviations from the $\Lambda$CDM model, but at the same time the error bars for the $\Lambda$CDM parameters become larger, in some cases a lot.

There are also a number of \emph{anomalies} in the WMAP data that cast some doubt on whether our current understanding of the universe is correct \cite{WMAP7anom}.  These are features that appear statistically unlikely in the scenario for the generation of primordial perturbations we have described. These anomalies include
\begin{itemize}
\item a low quadrupole: the observed $C_2$ is well below the expectation value of from the best-fit model (see Fig.~\ref{fig:wmap7TT})
\item ``axis of evil'': the pattern of the quadrupole and octupole parts of the CMB anisotropy are curiously aligned, having a common preferred direction \cite{Tegmark03}
\item cold spot: the cold region near the lower right edge in Fig.~\ref{fig:wmap7}
is unusually deep for such a small feature \cite{Vielva04}
\item north-south asymmetry: if one divides the celestial sphere along the ecliptic, 
the northern hemisphere (upper left in Fig.~\ref{fig:wmap7}) has much less large-scale anisotropy than the southern hemisphere (lower right) \cite{Eriksen04}
\end{itemize}

\section{Planck Mission}

\begin{figure}[ht]
\begin{center}
\includegraphics[width=10cm]{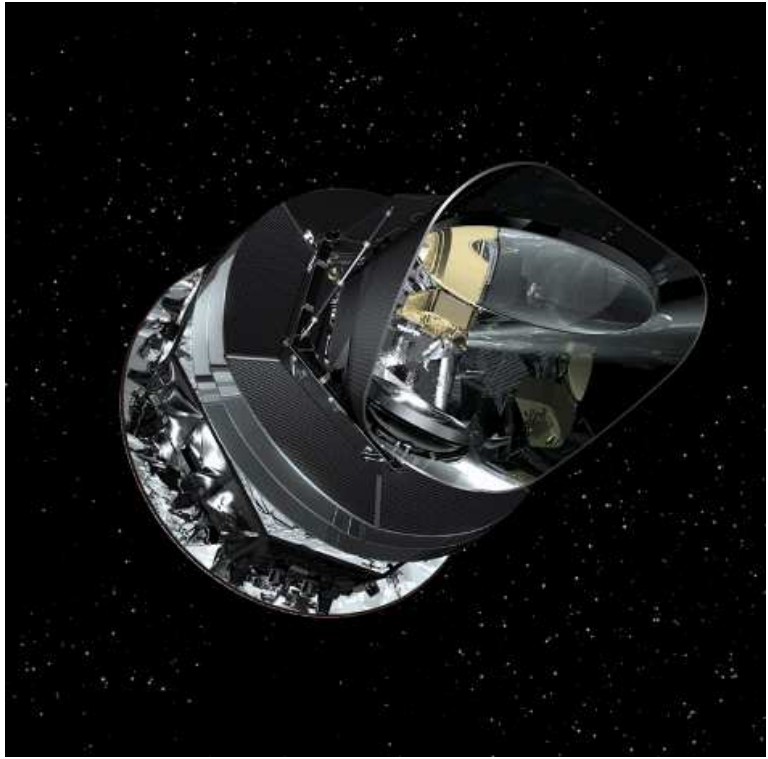}
\caption{An artist's view of Planck. Between the primary and secondary mirrors is the focal plane unit (see Fig.~\ref{fig:Planck_focal_plane}), to where the mirrors direct the microwave radiation.  These are protected from straylight by a baffle. Below and to the left of the baffle we see 
three layers of thermal shields that provide passive cooling. On the other side of the shields there is the warm service module.  The solar panels and the antenna for communication with ESA's ground stations are on its other side and are not visible in this view. From Ref.~\cite{Psite}.
 \emph{Credit: ESA/AOES Medialab.}}
\label{fig:PlanckFrontView}
\end{center}
\end{figure}

\begin{figure}[ht]
\begin{center}
\includegraphics[width=10cm]{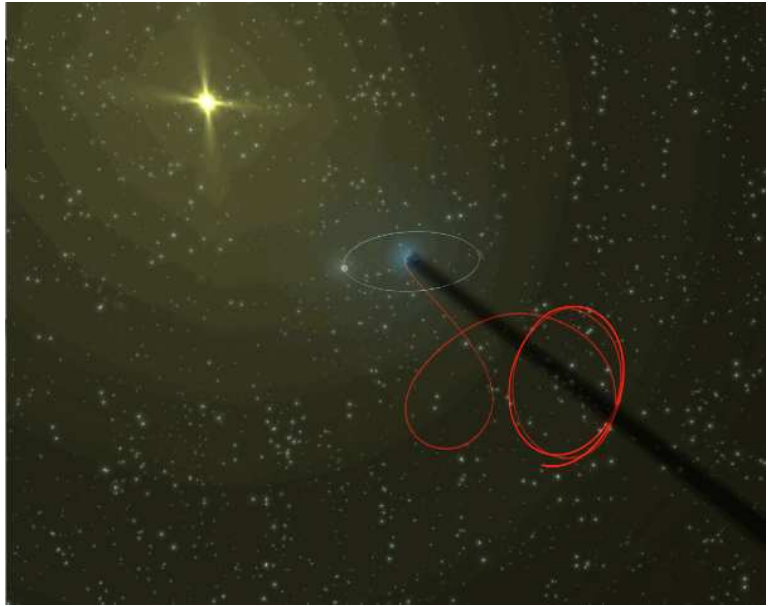}
\caption{Planck's cruise to and orbit around L2. From Ref.~\cite{Psite}.
 \emph{Credit: ESA - C. Carreau.}}
\label{fig:Planck_L2}
\end{center}
\end{figure}

\begin{figure}[ht]
\begin{center}
\includegraphics[width=9cm]{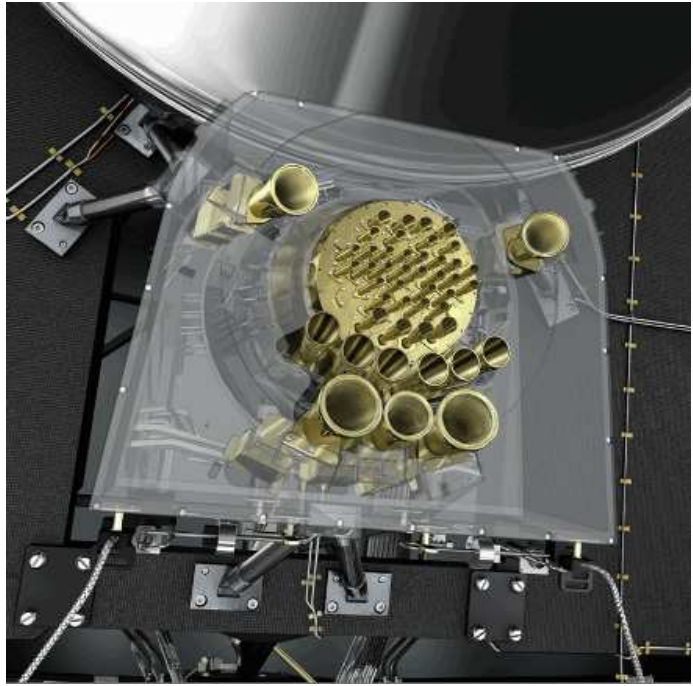}
\caption{The feedhorns of the Planck receivers at the focal plane of the Planck telescope. The smaller horns at the center belong to the HFI.  They are surrounded by the six 70 GHz, three 44 GHz, and two 30 GHz LFI feedhorns. From Ref.~\cite{Psite}.
 \emph{Credit: ESA/AOES Medialab.}}
\label{fig:Planck_focal_plane}
\end{center}
\end{figure}

\begin{figure}[ht]
\begin{center}
\includegraphics[width=16cm]{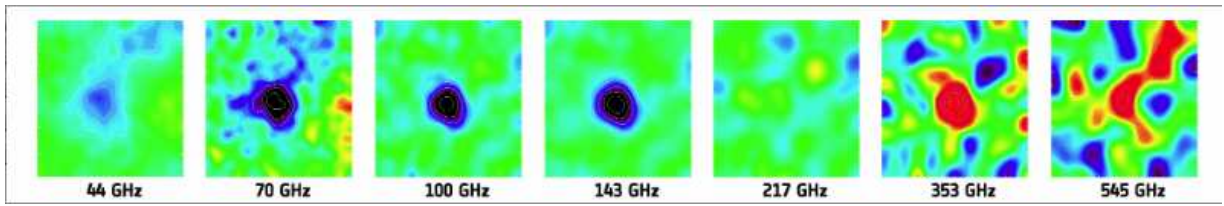}
\caption{Abell 2319, a nearby cluster of galaxies, seen by seven different Planck
channels. For frequencies below 217 GHz, the cluster appears as a cold spot, for frequencies above 217 GHz as a hot spot. From Ref.~\cite{Pscisite}. \emph{Credit: ESA/ LFI \& HFI Consortia.}}
\label{fig:SZ}
\end{center}
\end{figure}

\begin{figure}[ht]
\begin{center}
\includegraphics[width=16cm]{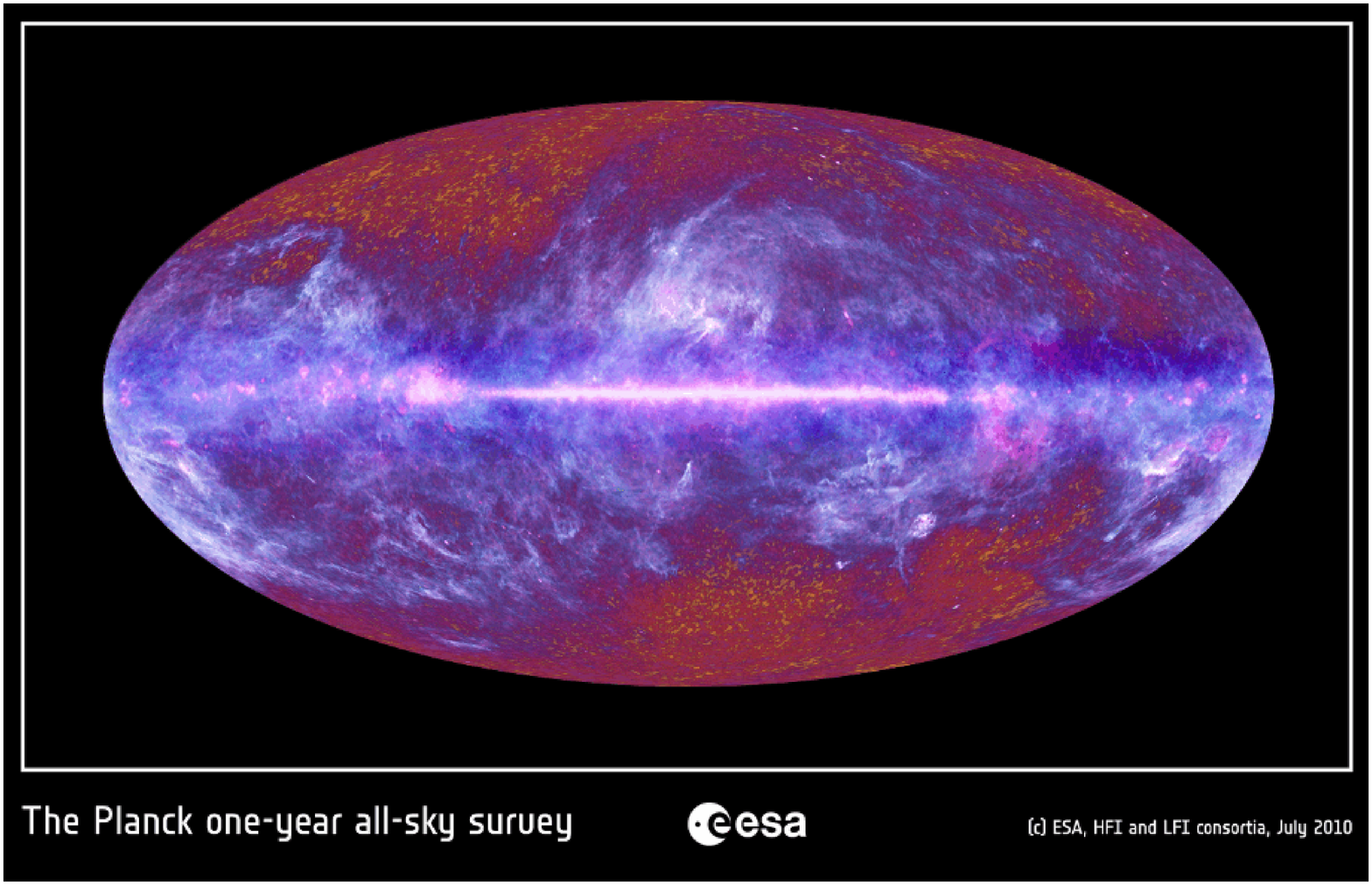}
\caption{A full-sky image of the microwave sky based on the first year of Planck observations. The colors in the image are related to the dominating microwave frequencies at the different parts of the sky. This image is made combining different Planck frequencies so that the foreground-dominated ones have been given the most weight, and therefore the CMB, shown in red and orange is only visible far from the galactic plane. In other parts of the sky the image mainly shows emission from the gas and dust in our own galaxy. From Ref.~\cite{Psite}. \emph{Credit: ESA/ LFI \& HFI Consortia.}}
\label{fig:Planck_full_sky}
\end{center}
\end{figure}

The next major improvement in our knowledge of the CMB anisotropy and polarization will come from the Planck satellite \cite{Pmission,Psite}, see Fig.~\ref{fig:PlanckFrontView}, which is a mission of the European Space Agency (ESA).

Planck was launched into space, together with the Herschel infrared space telescope, on an Ariane 5 rocket from  Europe's Spaceport in Kourou, French Guiana, on May 14th, 2009. Planck reached its orbit, around the 2nd Lagrange point (L2) of the Earth-Sun system, on July 2nd. See Fig.~\ref{fig:Planck_L2}.

Lagrange points are special points in celestial mechanics, where objects in free fall stay at the same relative position with respect to Earth and Sun. L2 is 1.5 million km from Earth in the anti-Sun direction.  L2 itself is in Earth's shadow, and since Planck draws its electric power from solar panels, it must not go too near L2.
Planck's orbit around L2 has a radius of 400 000 km. (WMAP is also in orbit around L2).

Compared to WMAP, Planck is an improvement in three respects:
\begin{itemize}
\item Planck has better angular resolution, 5 arcmin compared to WMAP's 14 arcmin
\item it has a wider frequency coverage, which is important for separating the CMB from foreground radiation
\item it has a higher sensitivity thanks to a much lower operating temperature
\end{itemize}

The higher angular resolution makes it possible to measure $C_\ell$ to much higher multipoles.  From Eq.~(\ref{eq:theta2ell}), while WMAP reached to $\ell \sim 800$, Planck reaches to $\ell \sim 2200$ (actually one gets information also from somewhat higher multipoles, but with less accuracy). This is especially important for measuring the spectral index to higher accuracy.

To cover the wider range of frequencies, Planck carries to separate instruments with two different detection technologies: The Low-Frequency Instrument (LFI) has radiometers for measurements at three frequency bands, centered at 30, 44, and 70 GHz.  The High-Frequency Instrument (HFI) has bolometers for measurements at six frequency bands, centered at 100, 143, 217, 353, 545, and 857 GHz. See Fig.~\ref{fig:Planck_focal_plane}.

The higher frequencies are important for separating out the microwave radiation from interstellar dust, whose radiation increases with frequency. They are also important for measuring the Sunyaev-Zeldovich (SZ) effect, which is the upscattering of CMB photons to higher frequencies by the energetic electrons in the intergalactic space of galaxy clusters.  In the direction of galaxy clusters, the SZ effect lowers the CMB intensity at frequencies below 217 GHz and raises it for frequencies above 217 GHz.  See Fig.~\ref{fig:SZ}. Although Planck is primarily a CMB mission, the full-sky observations at these many frequencies are also important for many astrophysical studies.

To achieve the high sensitivity of its instruments, Planck carries a three-stage active cooling system.  The first stage is a hydrogen sorption cooler to cool the LFI to 20 K. The second stage, the 4K cooler, is based on Joule-Thomson expansion of helium. In addition to being one stage of cooling the HFI, the 4K cooler also provides a reference load at 4 K temperature for the LFI radiometers. The final stage is a dilution cooler, which operates with helium-4 and helium-3, and is used to cool the HFI bolometers to 0.1 K.   Planck carries large tanks of helium-3 and helium-4, since after dilution these gases are vented to space. This limits the operating time of HFI, since this gas supply will eventually run out.  

The high sensitivity is especially important for polarization measurements, since the CMB polarization is at least an order of magnitude weaker than the temperature anisotropy. Except for the two highest frequencies, all Planck channels measure the polarization also.

The noise of the LFI radiometers is dominated by low noise frequencies. To remove this low-frequency noise the radiometers continuously observe the 4K reference load together with the sky. The signal from these two sources are switched at 8192 Hz frequency between the diodes of the radiometers, and are afterwards separated again into a sky signal and a reference signal. In this way both signal streams have come through the same electronics, and contain the same low-frequency noise. Taking the difference between the two signals removes most of the it, leaving the sky signal and residual noise that is almost white (uncorrelated).

Planck rotates at 1 rpm.  The instruments are pointed at a direction about $85^\circ$ away from the spin axis, scanning almost great circles on the sky. The spin axis is repointed by 2 arcmin about once per hour to keep it pointed close to the anti-Sun direction. To observe also the regions near the ecliptic poles the repointing scheme actually follows a cycloid around the anti-Sun direction so that the spin axis always points $7.5^\circ$ away from it. This way the whole sky is covered twice in one year.  Repeated measurements of the same sky points at different times is used at the map-making stage \cite{map-making} of the data analysis to remove residual correlated noise.

Planck reached the 0.1 K temperature about 50 days after launch, making the HFI bolometers the coldest known place in space! After that followed a period of tuning of the instruments and performance verification. On August 12th, science observations began with a two-week ``first-light survey'' (see Fig.~\ref{fig:fls}).  Without break the observations continued into the originally planned 15-month ``nominal survey''.
The whole sky was observed by June 2010.  See Fig.~\ref{fig:Planck_full_sky}. This nominal Planck mission ended on November 26th, 2010.  ESA has, however, extended Planck operations, both with LFI and HFI, by 12 months, until near the end of 2011.  LFI does not need the dilution cooler, and in November 2010 ESA extended the Planck observation program yet by another year, for using LFI only.

ESA has granted the Planck Collaboration a two-year proprietary period for data analysis and deriving science results, after which the data will be made public.  Thus the data from the nominal 15-month mission will be released near end of 2012, and the main cosmology results are expected to be published then also.  Some early results on the astrophysics of foregrounds will already be published in January 2011, including the \emph{Early Release Compact Source Catalog}.

\section{Cosmology from Planck}

\begin{figure}[ht]
\begin{center}
\includegraphics[width=12cm]{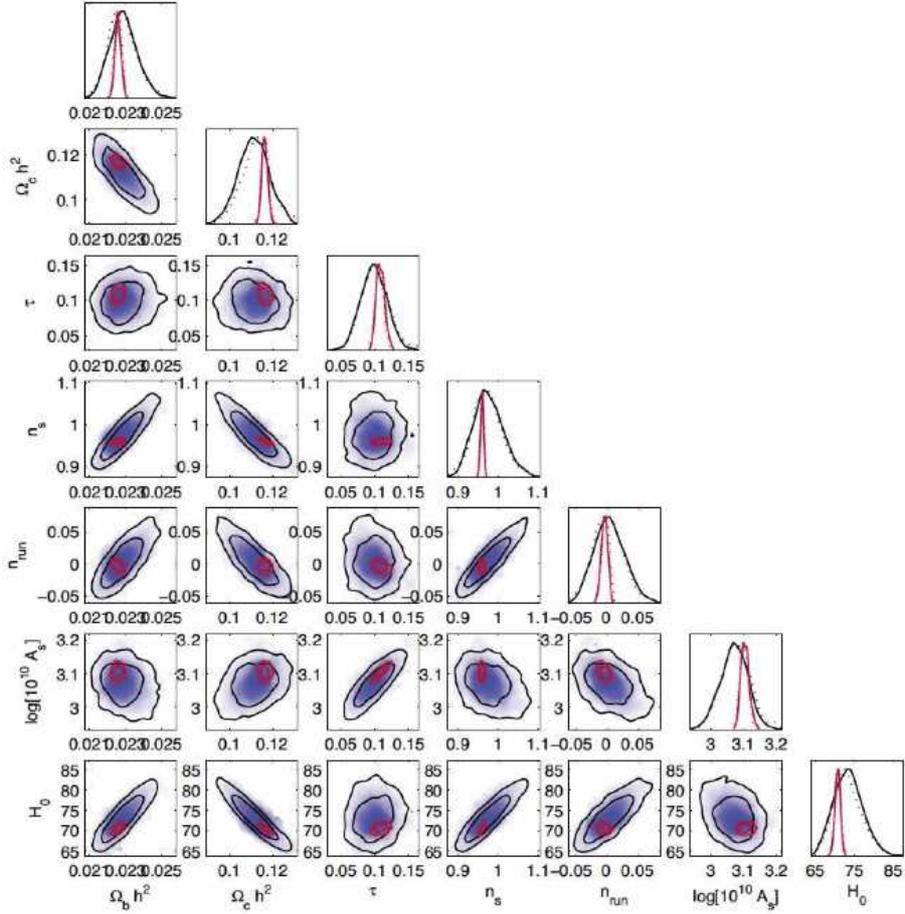}
\caption{An illustration of the expected improvement in cosmological parameter determination for a 7-parameter model: $\Lambda$CDM supplemented with an additional parameter $n_\mathrm{run} \equiv d n_s/d\ln k$, allowing the spectral index to vary with distance scale. Blue contours show forecasts for WMAP after 4 years of observation and red contours show results for Planck after 1 year of observations. From Ref.~\cite{bluebook}.}
\label{fig:P7params}
\end{center}
\end{figure}

Planck will extract almost all the available information from CMB temperature anisotropy, limited by cosmic variance and the ability to separate foregrounds from the CMB.  Planck will also provide an accurate spectrum of CMB polarization for the first time, although Planck is not optimized for polarization measurements, leaving room for a future CMB space mission focusing on polarization.

The higher sensitivity and resolution of Planck will lead to a major improvement in the accuracy of the determination of cosmological parameters. The main cosmological interest is in the nature of primordial perturbations. This is the key to the mechanism for the origin of structure in the universe, which is responsible for the existence of galaxies, stars, and planets. This mechanism is related to very high-energy physics whose study is beyond the reach of Earth-based accelerators.

A better determination of the spectral index $n_s$ of primordial perturbations is already important for selecting among candidate inflation theories. 
But even more important is the ability of Planck to probe additional cosmological parameters beyond the simple $\Lambda$CDM model (see Fig.~\ref{fig:P7params}). Many inflation models predict also the production of gravitational waves,
also called \emph{tensor perturbations}, in addition to the density (scalar) perturbations.  The upper limit from WMAP on the ratio of tensor perturbation amplitude to scalar perturbation amplitude is $r < 0.24$ (95\% CL) \cite{WMAP7cosmo}, assuming that this is the only extension to the $\Lambda$CDM model.
The effect of tensor perturbations on CMB temperature anisotropy is somewhat similar to some other cosmological parameters, but it causes an unambiguous signal in CMB polarization.

The polarization field on the celestial sphere can be divided into an E-mode (curl-free part) and a B-mode (source-free).  To first order in perturbation theory, scalar perturbations produce only E-mode polarization. Therefore a detection of B-mode polarization at relatively large scales $\ell < 150$ is  clear evidence of primordial tensor perturbations. At smaller scales, second-order effects convert a part of E-mode polarization into B-mode. The WMAP upper limit to tensor perturbations comes from CMB temperature anisotropy; the B-mode is beyond the reach of WMAP polarization sensitivity. Planck is sensitive enough to detect B-mode polarization coming from $r < 0.1$ \cite{bluebook}.

More complicated inflation models have additional signatures that are not included in the $\Lambda$CDM model, and Planck will be looking for these. They include deviations from the adiabaticity of primordial perturbations, and the deviation of their statistics from Gaussianity.

Another important thing expected from Planck is a check on the large-scale anomalies in WMAP data (Sec.~\ref{sec:wmap}).

\section*{Acknowledgements}

Planck is a project of the European Space Agency - ESA - with instruments provided by two scientific Consortia funded by ESA member states (in particular the lead countries: France and Italy) with contributions from NASA (USA), and telescope reflectors provided in a collaboration between ESA and a scientific Consortium led and funded by Denmark.
More information at {\tt http://www.esa.int/Planck}
The Finnish contribution is supported by the Finnish Funding Agency for Technology and Innovation (Tekes) and the Academy of Finland.

\section*{Bibliography}

V.F. Mukhanov, H.A. Feldman and R.H. Brandenberger,
Theory of Cosmological Perturbations,
\emph{Phys. Rep.} \textbf{215} (1992) 203. \\
A.R. Liddle and D.H. Lyth, \emph{Cosmological Inflation and Large-Scale Structure} 
(Cambridge University Press, Cambridge, 2000). \\
S. Dodelson, \emph{Modern Cosmology}
(Academic Press, 2003). \\
D.H. Lyth and A.R. Liddle, \emph{The Primordial Density Perturbation} (Cambridge University Press, Cambridge, 2009).


\begin{thebibliography}{99}


\bibitem{Psite} {\tt http://www.esa.int/Planck}

\bibitem{WMAPsite} {\tt http://map.gsfc.nasa.gov}

\bibitem{COBEtemp} D.J. Fixsen \emph{et al.}, \emph{Astrophys. J.} \textbf{473} (1996) 576; J.C. Mather \emph{et al.}, \emph{Astrophys. J.} \textbf{512} (1999) 511; 
D.J. Fixsen and J.C. Mather, \emph{Astrophys. J.} \textbf{581} (2002) 817.

\bibitem{PeWi} 
A.A. Wilson and R.W. Penzias, \emph{Astrophys. J.} \textbf{142} (1965) 419.

\bibitem{COBEanis} G.F. Smoot \emph{et al.}, \emph{Astrophys. J. Lett.} 
\textbf{396} (1992) L1.

\bibitem{lambda} {\tt http://lambda.gsfc.nasa.gov}

\bibitem{WMAP7spec} D. Larson \emph{et al.}, arXiv:1001.4635.

\bibitem{Keskitalo} R. Keskitalo, The effect of matter and baryon densities on the cosmic microwave background anisotropy, Master's thesis, University of Helsinki (2005).

\bibitem{WMAP7paper} N. Jarosik \emph{et al.}, arXiv:1001.4744.

\bibitem{WMAP7calib} J.L. Weiland \emph{et al.}, arXiv:1001.4731.

\bibitem{WMAP7cosmo} E. Komatsu \emph{et al.}, arXiv:1001.4538.


\bibitem{WMAP7anom} C.L. Bennett \emph{et al.}, arXiv:1001.4758.

\bibitem{Tegmark03} M. Tegmark, A. de Oliveira-Costa and A.J. Hamilton, 
\emph{Phys. Rev. D} \textbf{68} (2003) 123523.

\bibitem{Vielva04} P. Vielva \emph{et al.}, \emph{Astrophys. J.} 
\textbf{609} (2004) 22.

\bibitem{Eriksen04} H.K. Eriksen \emph{et al.}, \emph{Astrophys. J.} 
\textbf{605} (2004) 14.

\bibitem{Pmission} J. Tauber \emph{et al.}, \emph{Astron. Astrophys.} \textbf{520} (2010) A1.

\bibitem{Pscisite} {\tt http://sci.esa.int/planck}

\bibitem{map-making} H. Kurki-Suonio  \emph{et al.}, \emph{Astron. Astrophys.} \textbf{506} (2009) 1511; E. Keih\"{a}nen \emph{et al.}, \emph{Astron. Astrophys.} \textbf{510} (2010) 57.

\bibitem{bluebook} Planck Collaboration, The Scientific Programme of Planck (``Planck Bluebook''), ESA-SCI(2005)1, (astro-ph/0604069).

\end{thebibliography}
\end{document}